 \documentclass[pmlr,twocolumn]{jmlr} 

\usepackage{booktabs}
\usepackage[load-configurations=version-1]{siunitx} 

\usepackage{graphicx}
\theorembodyfont{\upshape}
\theoremheaderfont{\scshape}
\theorempostheader{:}
\theoremsep{\newline}

\jmlrvolume{ML4H Extended Abstract Arxiv Index}
\jmlryear{2020}
\jmlrsubmitted{2020}
\jmlrpublished{}
\jmlrworkshop{Machine Learning for Health (ML4H) 2020}

\title[CheXphotogenic: Generalization of Chest X-ray Models to  to Photos]{CheXphotogenic: Generalization of Deep Learning Models for Chest X-ray Interpretation to Photos of Chest X-rays}

\author{%
\Name{Pranav Rajpurkar}\thanks{Equal Contribution} \Email{pranavsr@cs.stanford.edu}\\
\Name{Anirudh Joshi}\footnotemark[1] \Email{anirudhjoshi@cs.stanford.edu}\\
\Name{Anuj Pareek}\footnotemark[1] \Email{anujpare@stanford.edu}\\
\Name{Jeremy Irvin} \Email{jirvin16@stanford.edu}\\
\Name{Andrew Y. Ng} \Email{ang@cs.stanford.edu}\\
\Name{Matthew Lungren} \Email{mlungren@cs.stanford.edu}}

\begin{document}

\maketitle

\begin{abstract}
The use of smartphones to take photographs of chest x-rays represents an appealing solution for scaled deployment of deep learning models for chest x-ray interpretation.
However, the performance of chest x-ray algorithms on photos of chest x-rays has not been thoroughly investigated.
In this study, we measured the diagnostic performance for 8 different chest x-ray models when applied to photos of chest x-rays.
All models were developed by different groups and submitted to the CheXpert challenge, and re-applied to smartphone photos of x-rays in the CheXphoto dataset without further tuning.
We found that several models had a drop in performance when applied to photos of chest x-rays, but even with this drop, some models still performed comparably to radiologists.
Further investigation could be directed towards understanding how different model training procedures may affect model generalization to photos of chest x-rays.

\end{abstract}

\section{Introduction}
\label{sec:intro}

Chest x-rays are the most common imaging examination in the world, critical for diagnosis and management of many diseases. With over 2 billion chest x-rays performed globally annually, many clinics in both developing and developed countries lack sufficient trained radiologists to perform timely x-ray interpretation. Automating cognitive tasks in medical imaging interpretation with deep learning models could improve access, efficiency, and augment existing workflows \citep{rajpurkar_deep_2018, nam_development_2018, singh_deep_2018, qin_computer-aided_2018}. However, a major obstacle to clinical adoption of such technologies is in model deployment, an effort often frustrated by vast heterogeneity of clinical workflows across the world \citep{kelly_key_2019}. Chest x-ray models are developed and validated using digital x-rays with many deployment solutions relying on heavily integrated yet often disparate infrastructures \citep{qin_using_2019, lakhani_deep_2017, kallianos_how_2019, kashyap_artificial_2019, shih_augmenting_2019}.

\begin{figure}[t]
\centering
  \includegraphics[width=\columnwidth]{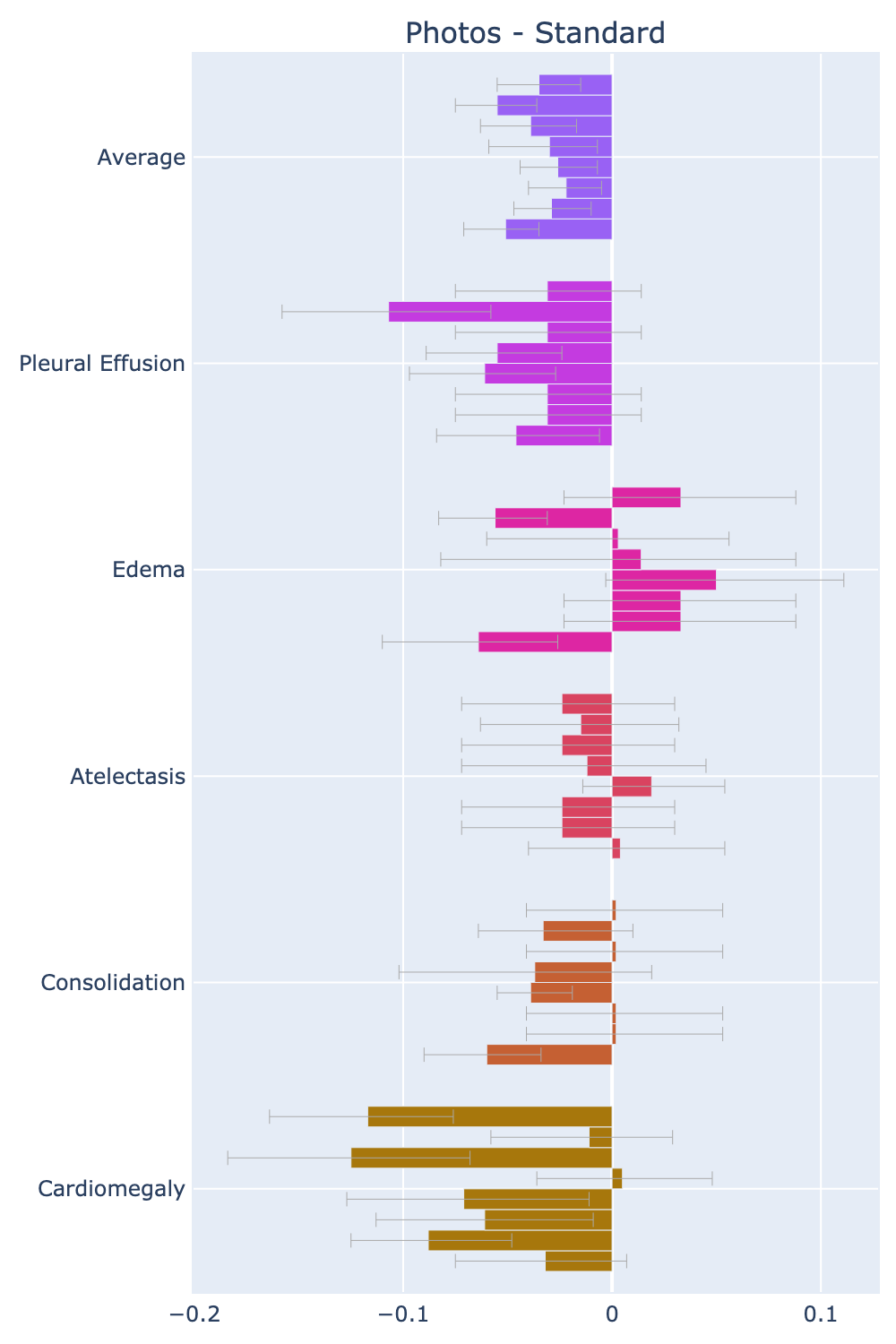}
  \caption{MCC differences of 8 chest x-ray models on different pathologies between photos of the x-rays and the original x-rays with 95\% confidence intervals.}
\label{fig:standard}
\end{figure}

\begin{figure}[t]
\centering
  \includegraphics[width=\columnwidth]{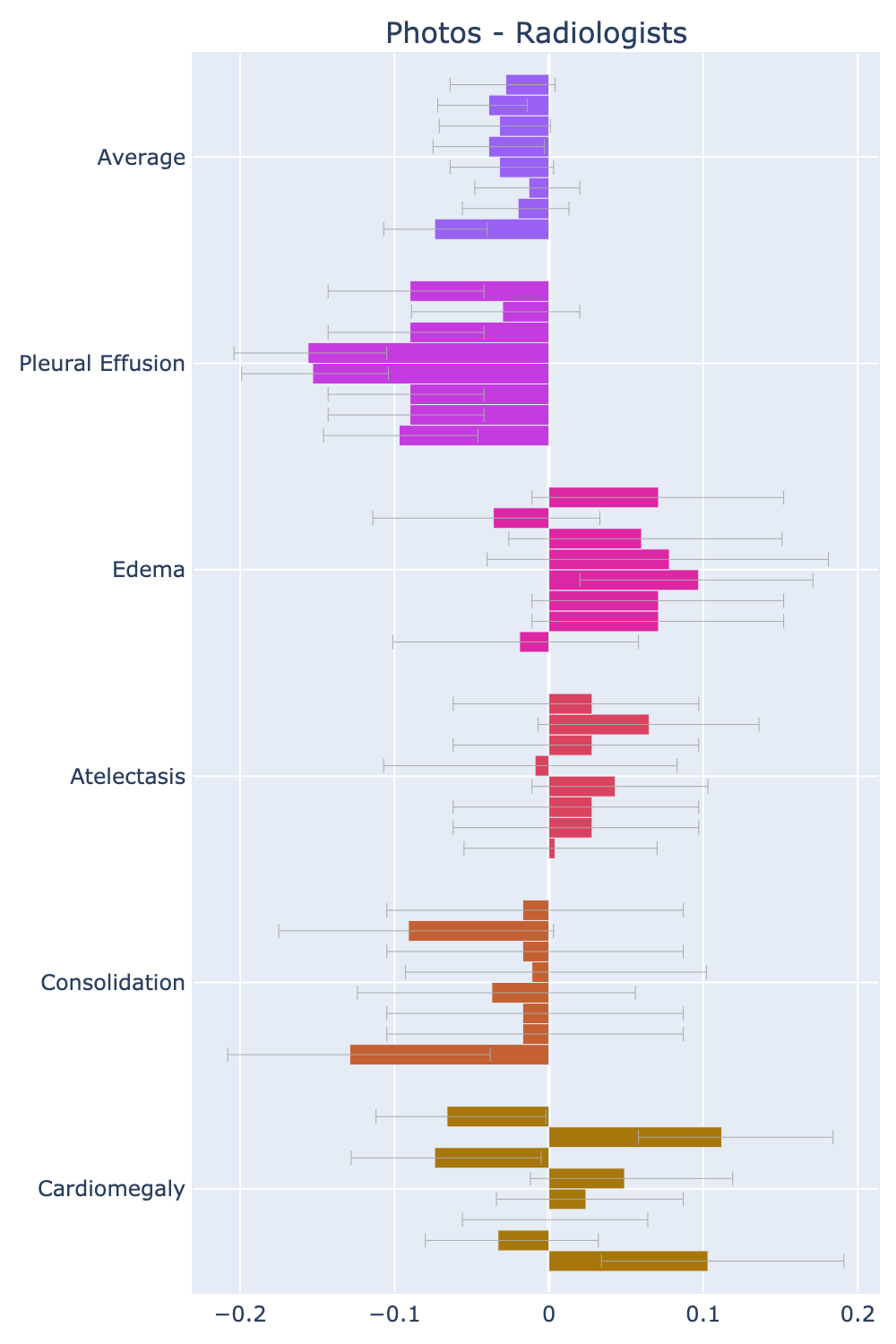}
  \caption{MCC differences of the same models on photos of chest x-rays compared to radiologist performance with 95\% confidence intervals. }
  \label{fig:photos}
\end{figure}

One appealing solution to scaled deployment across disparate clinical frameworks is to leverage the ubiquity of smartphones.  Interpretation of medical imaging via cell phone photography is an existing ``store-and-forward telemedicine'' approach in which one or more photos of medical imaging are captured and sent as email attachments or instant messages by practitioners to obtain second opinions from specialists in routine clinical care \citep{goost2012image, vassallo1998first}. 
Smartphone photographs have been shown to be of sufficient diagnostic quality to allow for medical interpretation, thus leveraging deep learning models in automated interpretation of photos of medical imaging examinations may serve as an infrastructure agnostic approach to deployment, particularly in resource limited settings. However, significant technical barriers exist in automated interpretation of photos of chest x-rays. Photographs of x-rays introduce visual artifacts which are not commonly found in digital x-rays, such as altered viewing angles, variable lighting conditions, glare, moiré, rotations, translations, and blur \citep{phillips2020chexphoto}. These artifacts have been shown to reduce algorithm performance when input images are perceived through a camera \citep{kurakin_adversarial_2016}. 

We measured the diagnostic performance for 8 different chest x-ray models when applied to photos of chest x-rays. All models were developed by different groups and submitted to the CheXpert challenge, a large public competition for digital chest x-ray analysis \citep{irvin_chexpert_2019}. We applied these models to a dataset of smartphone photos of 668 x-rays from 500 patients. Models were evaluated on their diagnostic performance in binary classification, as measured by Matthew’s Correlation Coefficient (MCC) \citep{chicco2020advantages}, on the following pathologies selected in \citet{irvin_chexpert_2019}: atelectasis, cardiomegaly, consolidation, edema, and pleural effusion \citep{irvin_chexpert_2019}. 

We found that several chest x-ray models had a drop in performance when applied to smartphone photos of chest x-rays, but even with this drop, some models still performed comparably to radiologists.


\section{Methods} 


\paragraph{Models.} We investigated the generalization performance of 8 available models on the CheXpert competition \citep{irvin_chexpert_2019}. CheXpert is a competition for automated chest x-ray interpretation that has been running since January 2019 featuring a strong radiologist-labeled reference standard. Many models have been submitted to the CheXpert leaderboard from both academic and industry teams. The top 8 available models of the 94 models on the CheXpert competition leaderboard as of November 2019 were selected. All of the selected models were ensembles with the number of models in the ensemble ranging from 8 to 32; the majority of these models featured Densely Connected Convolutional Networks \citep{huang2017densely}.  
 
\paragraph{Test Set.} CheXpert used a hidden test set for official evaluation of models. Teams submitted their executable code, which was then run on a test set that was not publicly readable to preserve the integrity of the test results. We made use of the CodaLab platform to re-run these chest x-ray models.
We evaluated these models on the CheXphoto \citep{phillips2020chexphoto} test set, a dataset of photos of the x-rays from the CheXpert test set.


\begin{table}[h]
    \resizebox{\columnwidth}{!}{  
    \begin{tabular}{c|l|c}
 \hline
 & Comparison & Result  \\ 
 \hline
AUC & Photos & 0.856 (0.840, 0.869) \\ 
 & Standard & 0.871 (0.855, 0.863) \\ 
 \hline
 AUC & Standard-Photos & 0.016 (0.012, 0.019) \\
 \hline
MCC & Photos & 0.560 (0.528, 0.587) \\
& Standard & 0.588 (0.560, 0.618)  \\
& Radiologists & 0.568 (0.542, 0.597) \\
\hline
MCC & Standard-Photos & 0.029 (0.014, 0.043) \\
& Radiologists-Photos & 0.009 (-0.022, 0.042)\\
 \hline
 \end{tabular}
 }
  \caption{AUC and MCC performance of models and radiologists on the standard x-rays and the photos of chest x-rays, with 95\% confidence intervals.}
\label{tab:tab1}
\end{table}

\paragraph{Evaluation Metrics.}
Our primary evaluation metric was Matthew’s Correlation Coefficient (MCC), a statistical rate which produces a high score only if the prediction obtained good results in all of the four confusion matrix categories (true positives, false negatives, true negatives, and false positives); MCC is proportionally both to the size of positive elements and the size of negative elements in the dataset \citep{chicco2020advantages}.

We reported the average MCC of 8 models for five pathologies, namely atelectasis, cardiomegaly, consolidation, edema, and pleural effusion. Additionally, in experiments comparing the models on standard chest x-rays to photos of chest x-rays, we reported the AUC and MCC of the models. In experiments comparing models to board-certified radiologists, we reported the difference in MCC for each of the five pathologies.

\section{Results}
In comparison of model performance on digital chest x-rays to photos, all eight models experienced a statistically significant drop in task performance on photos with an average drop of 0.036 MCC (95\% CI 0.024, 0.048) (See Figure \ref{fig:standard}, Table \ref{tab:tab1}). All models had a statistically significant drop on at least one of the pathologies between native digital image to photos. One model had a statistically significant drop in performance on three pathologies: pleural effusion, edema, and consolidation. Two models had a significant drop on two pathologies: one on pleural effusion and edema, and the other on pleural effusion and cardiomegaly. The cardiomegaly and pleural effusion tasks led to decreased performance in five and four models respectively.



In comparison of performance of models on photos compared to radiologist performance, three out of eight models performed significantly worse than radiologists on average, and the other five had no significant difference (see Figure \ref{fig:photos}). On specific pathologies, there were some models that had a significantly higher performance than radiologists: two models on cardiomegaly, and one model on edema. Conversely, there were some models that had a significantly lower performance than radiologists: two models on cardiomegaly, and one model on consolidation.  The pathology with the greatest number of models that had a significantly lower performance than radiologists was pleural effusion (seven models).

\section{Discussion} 
 
Our results demonstrated that while most models experienced a significant drop in performance when applied to photos of chest x-rays compared to the native digital image, their performance was nonetheless largely equivalent to radiologist performance. We found that although there were thirteen times that models had a statistically significant drop in performance on photos on the different pathologies, the models had significantly lower performance than radiologists only 6 of those 13 times. Comparison to radiologist performance provides context in regard to clinical applicability: several models remained comparable to radiologist performance standard despite decreased performance on photos.

While using photos of chest x-rays to input into chest x-ray algorithms could enable any physician with a smartphone to get instant AI algorithm assistance, the performance of chest x-ray algorithms on photos of chest x-rays has not been thoroughly investigated. Several studies have highlighted the importance of generalizability of computer vision models with noise in  \citep{hendrycks_benchmarking_2019}. \citet{dodge_study_2017} demonstrated that deep neural networks perform poorly compared to humans on image classification on distorted images. \citet{geirhos_imagenet-trained_2019}, \citet{schmidt_adversarially_2018} have found that convolutional neural networks trained on specific image corruptions did not generalize, and the error patterns of network and human predictions were not similar on noisy and elastically deformed images. Our work makes significant contributions over another investigation of chest x-ray models \citep{rajpurkar2020chexpedition}. While their study considered the differences in AUC of models when applied to photos of x-rays, they did not (1) compare the resulting performances against radiologists, (2) investigate the drop in performances on specific tasks, or (3) analyze drops in performances of individual models across tasks. 

Further investigation could be directed towards understanding how different model training procedures may affect model generalization to photos of chest x-rays, and understanding etiologies behind trends for changes in performance for specific pathologies or specific artifacts.

\bibliography{jmlr-sample}

\begin{thebibliography}{22}
\providecommand{\natexlab}[1]{#1}
\providecommand{\url}[1]{\texttt{#1}}
\expandafter\ifx\csname urlstyle\endcsname\relax
  \providecommand{\doi}[1]{doi: #1}\else
  \providecommand{\doi}{doi: \begingroup \urlstyle{rm}\Url}\fi

\bibitem[Chicco and Jurman(2020)]{chicco2020advantages}
Davide Chicco and Giuseppe Jurman.
\newblock The advantages of the matthews correlation coefficient (mcc) over f1
  score and accuracy in binary classification evaluation.
\newblock \emph{BMC genomics}, 21\penalty0 (1):\penalty0 6, 2020.

\bibitem[Dodge and Karam(2017)]{dodge_study_2017}
Samuel Dodge and Lina Karam.
\newblock A {Study} and {Comparison} of {Human} and {Deep} {Learning}
  {Recognition} {Performance} under {Visual} {Distortions}.
\newblock In \emph{2017 26th {International} {Conference} on {Computer}
  {Communication} and {Networks} ({ICCCN})}, pages 1--7, July 2017.
\newblock \doi{10.1109/ICCCN.2017.8038465}.

\bibitem[Geirhos et~al.(2019)Geirhos, Rubisch, Michaelis, Bethge, Wichmann, and
  Brendel]{geirhos_imagenet-trained_2019}
Robert Geirhos, Patricia Rubisch, Claudio Michaelis, Matthias Bethge, Felix~A.
  Wichmann, and Wieland Brendel.
\newblock {ImageNet}-trained {CNNs} are biased towards texture; increasing
  shape bias improves accuracy and robustness.
\newblock \emph{arXiv:1811.12231 [cs, q-bio, stat]}, January 2019.

\bibitem[Goost et~al.(2012)Goost, Witten, Heck, Hadizadeh, Weber, Gr{\"a}ff,
  Burger, Montag, Koerfer, and Kabir]{goost2012image}
Hans Goost, Johannes Witten, Andreas Heck, Dariusch~R Hadizadeh, Oliver Weber,
  Ingo Gr{\"a}ff, Christof Burger, Mareen Montag, Felix Koerfer, and Koroush
  Kabir.
\newblock Image and diagnosis quality of x-ray image transmission via cell
  phone camera: a project study evaluating quality and reliability.
\newblock \emph{PLoS One}, 7\penalty0 (10):\penalty0 e43402, 2012.

\bibitem[Hendrycks and Dietterich(2019)]{hendrycks_benchmarking_2019}
Dan Hendrycks and Thomas Dietterich.
\newblock Benchmarking {Neural} {Network} {Robustness} to {Common}
  {Corruptions} and {Perturbations}.
\newblock \emph{arXiv:1903.12261 [cs, stat]}, March 2019.

\bibitem[Huang et~al.(2017)Huang, Liu, Van Der~Maaten, and
  Weinberger]{huang2017densely}
Gao Huang, Zhuang Liu, Laurens Van Der~Maaten, and Kilian~Q Weinberger.
\newblock Densely connected convolutional networks.
\newblock In \emph{Proceedings of the IEEE conference on computer vision and
  pattern recognition}, pages 4700--4708, 2017.

\bibitem[Irvin et~al.(2019)Irvin, Rajpurkar, Ko, Yu, Ciurea-Ilcus, Chute,
  Marklund, Haghgoo, Ball, Shpanskaya, Seekins, Mong, Halabi, Sandberg, Jones,
  Larson, Langlotz, Patel, Lungren, and Ng]{irvin_chexpert_2019}
Jeremy Irvin, Pranav Rajpurkar, Michael Ko, Yifan Yu, Silviana Ciurea-Ilcus,
  Chris Chute, Henrik Marklund, Behzad Haghgoo, Robyn Ball, Katie Shpanskaya,
  Jayne Seekins, David~A. Mong, Safwan~S. Halabi, Jesse~K. Sandberg, Ricky
  Jones, David~B. Larson, Curtis~P. Langlotz, Bhavik~N. Patel, Matthew~P.
  Lungren, and Andrew~Y. Ng.
\newblock {CheXpert}: {A} {Large} {Chest} {Radiograph} {Dataset} with
  {Uncertainty} {Labels} and {Expert} {Comparison}.
\newblock \emph{Proceedings of the AAAI Conference on Artificial Intelligence},
  33:\penalty0 590--597, July 2019.
\newblock ISSN 2374-3468, 2159-5399.
\newblock \doi{10.1609/aaai.v33i01.3301590}.
\newblock URL \url{https://aaai.org/ojs/index.php/AAAI/article/view/3834}.

\bibitem[Kallianos et~al.(2019)Kallianos, Mongan, Antani, Henry, Taylor, Abuya,
  and Kohli]{kallianos_how_2019}
K.~Kallianos, J.~Mongan, S.~Antani, T.~Henry, A.~Taylor, J.~Abuya, and
  M.~Kohli.
\newblock How far have we come? {Artificial} intelligence for chest radiograph
  interpretation.
\newblock \emph{Clinical Radiology}, 74\penalty0 (5):\penalty0 338--345, May
  2019.
\newblock ISSN 0009-9260.
\newblock \doi{10.1016/j.crad.2018.12.015}.

\bibitem[Kashyap et~al.(2019)Kashyap, Moradi, Karargyris, Wu, Morris, Saboury,
  Siegel, and Syeda-Mahmood]{kashyap_artificial_2019}
Satyananda Kashyap, Mehdi Moradi, Alexandros Karargyris, Joy~T. Wu, Michael
  Morris, Babak Saboury, Eliot Siegel, and Tanveer Syeda-Mahmood.
\newblock Artificial intelligence for point of care radiograph quality
  assessment.
\newblock In \emph{Medical {Imaging} 2019: {Computer}-{Aided} {Diagnosis}},
  volume 10950, page 109503K. International Society for Optics and Photonics,
  March 2019.
\newblock \doi{10.1117/12.2513092}.

\bibitem[Kelly et~al.(2019)Kelly, Karthikesalingam, Suleyman, Corrado, and
  King]{kelly_key_2019}
Christopher~J. Kelly, Alan Karthikesalingam, Mustafa Suleyman, Greg Corrado,
  and Dominic King.
\newblock Key challenges for delivering clinical impact with artificial
  intelligence.
\newblock \emph{BMC Medicine}, 17\penalty0 (1):\penalty0 195, December 2019.
\newblock ISSN 1741-7015.
\newblock \doi{10.1186/s12916-019-1426-2}.
\newblock URL
  \url{https://bmcmedicine.biomedcentral.com/articles/10.1186/s12916-019-1426-2}.

\bibitem[Kurakin et~al.(2016)Kurakin, Goodfellow, and
  Bengio]{kurakin_adversarial_2016}
Alexey Kurakin, Ian~J. Goodfellow, and Samy Bengio.
\newblock Adversarial examples in the physical world.
\newblock \emph{CoRR}, abs/1607.02533, 2016.
\newblock \_eprint: 1607.02533.

\bibitem[Lakhani and Sundaram(2017)]{lakhani_deep_2017}
Paras Lakhani and Baskaran Sundaram.
\newblock Deep {Learning} at {Chest} {Radiography}: {Automated}
  {Classification} of {Pulmonary} {Tuberculosis} by {Using} {Convolutional}
  {Neural} {Networks}.
\newblock \emph{Radiology}, 284\penalty0 (2):\penalty0 574--582, April 2017.
\newblock ISSN 0033-8419.
\newblock \doi{10.1148/radiol.2017162326}.

\bibitem[Nam et~al.(2018)Nam, Park, Hwang, Lee, Jin, Lim, Vu, Sohn, Hwang, Goo,
  and {others}]{nam_development_2018}
Ju~Gang Nam, Sunggyun Park, Eui~Jin Hwang, Jong~Hyuk Lee, Kwang-Nam Jin,
  Kun~Young Lim, Thienkai~Huy Vu, Jae~Ho Sohn, Sangheum Hwang, Jin~Mo Goo, and
  {others}.
\newblock Development and validation of deep learning–based automatic
  detection algorithm for malignant pulmonary nodules on chest radiographs.
\newblock \emph{Radiology}, 290\penalty0 (1):\penalty0 218--228, 2018.
\newblock Publisher: Radiological Society of North America.

\bibitem[Phillips et~al.(2020)Phillips, Rajpurkar, Sabini, Krishnan, Zhou,
  Pareek, Phu, Wang, Ng, and Lungren]{phillips2020chexphoto}
Nick~A. Phillips, Pranav Rajpurkar, Mark Sabini, Rayan Krishnan, Sharon Zhou,
  Anuj Pareek, Nguyet~Minh Phu, Chris Wang, Andrew~Y. Ng, and Matthew~P.
  Lungren.
\newblock Chexphoto: 10,000+ smartphone photos and synthetic photographic
  transformations of chest x-rays for benchmarking deep learning robustness,
  2020.

\bibitem[Qin et~al.(2018)Qin, Yao, Shi, and Song]{qin_computer-aided_2018}
Chunli Qin, Demin Yao, Yonghong Shi, and Zhijian Song.
\newblock Computer-aided detection in chest radiography based on artificial
  intelligence: a survey.
\newblock \emph{BioMedical Engineering OnLine}, 17\penalty0 (1):\penalty0 113,
  August 2018.
\newblock ISSN 1475-925X.
\newblock \doi{10.1186/s12938-018-0544-y}.

\bibitem[Qin et~al.(2019)Qin, Sander, Rai, Titahong, Sudrungrot, Laah,
  Adhikari, Carter, Puri, Codlin, and Creswell]{qin_using_2019}
Zhi~Zhen Qin, Melissa~S. Sander, Bishwa Rai, Collins~N. Titahong, Santat
  Sudrungrot, Sylvain~N. Laah, Lal~Mani Adhikari, E.~Jane Carter, Lekha Puri,
  Andrew~J. Codlin, and Jacob Creswell.
\newblock Using artificial intelligence to read chest radiographs for
  tuberculosis detection: {A} multi-site evaluation of the diagnostic accuracy
  of three deep learning systems.
\newblock \emph{Scientific Reports}, 9\penalty0 (1):\penalty0 1--10, October
  2019.
\newblock ISSN 2045-2322.
\newblock \doi{10.1038/s41598-019-51503-3}.

\bibitem[Rajpurkar et~al.(2018)Rajpurkar, Irvin, Ball, Zhu, Yang, Mehta, Duan,
  Ding, Bagul, Langlotz, Patel, Yeom, Shpanskaya, Blankenberg, Seekins,
  Amrhein, Mong, Halabi, Zucker, Ng, and Lungren]{rajpurkar_deep_2018}
Pranav Rajpurkar, Jeremy Irvin, Robyn~L. Ball, Kaylie Zhu, Brandon Yang,
  Hershel Mehta, Tony Duan, Daisy Ding, Aarti Bagul, Curtis~P. Langlotz,
  Bhavik~N. Patel, Kristen~W. Yeom, Katie Shpanskaya, Francis~G. Blankenberg,
  Jayne Seekins, Timothy~J. Amrhein, David~A. Mong, Safwan~S. Halabi, Evan~J.
  Zucker, Andrew~Y. Ng, and Matthew~P. Lungren.
\newblock Deep learning for chest radiograph diagnosis: {A} retrospective
  comparison of the {CheXNeXt} algorithm to practicing radiologists.
\newblock \emph{PLOS Medicine}, 15\penalty0 (11):\penalty0 e1002686, November
  2018.
\newblock ISSN 1549-1676.
\newblock \doi{10.1371/journal.pmed.1002686}.

\bibitem[Rajpurkar et~al.(2020)Rajpurkar, Joshi, Pareek, Chen, Kiani, Irvin,
  Ng, and Lungren]{rajpurkar2020chexpedition}
Pranav Rajpurkar, Anirudh Joshi, Anuj Pareek, Phil Chen, Amirhossein Kiani,
  Jeremy Irvin, Andrew~Y. Ng, and Matthew~P. Lungren.
\newblock Chexpedition: Investigating generalization challenges for translation
  of chest x-ray algorithms to the clinical setting, 2020.

\bibitem[Schmidt et~al.(2018)Schmidt, Santurkar, Tsipras, Talwar, and
  Madry]{schmidt_adversarially_2018}
Ludwig Schmidt, Shibani Santurkar, Dimitris Tsipras, Kunal Talwar, and
  Aleksander Madry.
\newblock Adversarially {Robust} {Generalization} {Requires} {More} {Data}.
\newblock In S.~Bengio, H.~Wallach, H.~Larochelle, K.~Grauman, N.~Cesa-Bianchi,
  and R.~Garnett, editors, \emph{Advances in {Neural} {Information}
  {Processing} {Systems} 31}, pages 5014--5026. Curran Associates, Inc., 2018.

\bibitem[Shih et~al.(2019)Shih, Wu, Halabi, Kohli, Prevedello, Cook, Sharma,
  Amorosa, Arteaga, Galperin-Aizenberg, Gill, Godoy, Hobbs, Jeudy, Laroia,
  Shah, Vummidi, Yaddanapudi, and Stein]{shih_augmenting_2019}
George Shih, Carol~C. Wu, Safwan~S. Halabi, Marc~D. Kohli, Luciano~M.
  Prevedello, Tessa~S. Cook, Arjun Sharma, Judith~K. Amorosa, Veronica Arteaga,
  Maya Galperin-Aizenberg, Ritu~R. Gill, Myrna~C.B. Godoy, Stephen Hobbs, Jean
  Jeudy, Archana Laroia, Palmi~N. Shah, Dharshan Vummidi, Kavitha Yaddanapudi,
  and Anouk Stein.
\newblock Augmenting the {National} {Institutes} of {Health} {Chest}
  {Radiograph} {Dataset} with {Expert} {Annotations} of {Possible} {Pneumonia}.
\newblock \emph{Radiology: Artificial Intelligence}, 1\penalty0 (1):\penalty0
  e180041, January 2019.
\newblock \doi{10.1148/ryai.2019180041}.

\bibitem[Singh et~al.(2018)Singh, Kalra, Nitiwarangkul, Patti, Homayounieh,
  Padole, Rao, Putha, Muse, Sharma, and Digumarthy]{singh_deep_2018}
Ramandeep Singh, Mannudeep~K. Kalra, Chayanin Nitiwarangkul, John~A. Patti,
  Fatemeh Homayounieh, Atul Padole, Pooja Rao, Preetham Putha, Victorine~V.
  Muse, Amita Sharma, and Subba~R. Digumarthy.
\newblock Deep learning in chest radiography: {Detection} of findings and
  presence of change.
\newblock \emph{PLoS ONE}, 13\penalty0 (10), October 2018.
\newblock ISSN 1932-6203.
\newblock \doi{10.1371/journal.pone.0204155}.

\bibitem[Vassallo et~al.(1998)Vassallo, Buxton, Kilbey, and
  Trasler]{vassallo1998first}
DJ~Vassallo, PJ~Buxton, JH~Kilbey, and M~Trasler.
\newblock The first telemedicine link for the british forces.
\newblock \emph{Journal of the Royal Army Medical Corps}, 144\penalty0
  (3):\penalty0 125--130, 1998.

\end{thebibliography}






\end{document}